\documentclass[12pt]{article}

\usepackage{graphicx}
\usepackage{hyperref}
\usepackage{mathrsfs}
\usepackage{latexsym}
\usepackage{amsfonts}
\usepackage{amsmath}
\usepackage{esint}
\usepackage{amsxtra}

\def \m{\mbox}
\def \be{\begin{equation}}
\def \ee{\end{equation}}
\def\beq{\begin{eqnarray}}
\def\eeq{\end{eqnarray}}
\def \ba{\begin{array}}
\def \ea{\end{array}}
\def\sin{\mbox{sin}}

\def\nn{\nonumber}

\def \f{\frac}

\def \p{\partial}

\begin{document}
\vspace*{-.6in}
\thispagestyle{empty}
\baselineskip = 18pt

\vspace{.5in}
\vspace{.5in}
{\LARGE
\begin{center}
Matone's relation of $\mathcal{N}=2$ super Yang-Mills\\ and spectrum
of Toda chain
\end{center}}

\vspace{1.0cm}

\begin{center}

Wei He\footnote{weihe@itp.ac.cn} \\
\vspace{1.0cm}\emph{Center of Mathematical Sciences, Zhejiang
University, Hangzhou 310027, China}
\end{center}
\vspace{1.0cm}

\begin{center}
\textbf{Abstract}
\end{center}
\begin{quotation}
\noindent In $\mathcal {N}=2$ super Yang-Mills theory, the Matone's
relation relates instanton corrections of the prepotential to
instanton corrections of scalar field condensation
$<\m{Tr}\varphi^2>$. This relation has been proved to hold for Omega
deformed theories too, using localization method. In this paper, we
first give a case study supporting the relation, which does not rely
on the localization technique. Especially, we show that the magnetic
expansion also satisfies a relation of Matone's type. Then we
discuss implication of the relation for the spectrum of periodic
Toda chain, in the context of recently proposed Nekrasov-Shatashvili
scheme.
\end{quotation}

\pagenumbering{arabic}

\newpage

\section{Introduction}

The $\mathcal {N}=2$ super Yang-Mills theory can be analytically
solved along the line of Seiberg-Witten theory of SU(2) gauge
theory\cite{SW9407}. The low energy dynamics of Coulomb phase of
gauge theory is encoded in the prepotential $\mathcal {F}(a)$, a
holomorphic function of the complex quantity $a$. And
$\f{1}{2}a\sigma_3$ is the vacuum expectation value(v.e.v) of the
complex scalar field $\varphi$, $u(a)=<\m{Tr}\varphi^2>$ is the
proper parameter for the moduli space. The prepotential receives
contributions from perturbative(tree and 1-loop) effects and
nonperturbative effects,\be \mathcal {F}(a)=\mathcal
{F}^{pert}(a)+\mathcal {F}^{inst}(a).\ee They have the form of power
expansion as \be \mathcal
{F}^{pert}(a)=\f{i}{\pi}a^2(\m{ln}(\f{2a}{\Lambda})-\f{3}{2}),
\qquad \mathcal
{F}^{inst}(a)=\sum_{k=1}^{\infty}\f{i}{\pi}c_ka^2(\f{\Lambda}{a})^{4k}.\ee
where $\Lambda$ is the scale of gauge theory.  $\mathcal{F}(a)$ and
$u(a)$ satisfy the following remarkable relation, \be
2\mathcal{F}(a)-a\f{\p\mathcal{F}(a)}{\p
a}=\f{2u}{i\pi}.\label{elecrelat}\ee

The most interesting, and nontrival, part of the prepotential is the
noperturbative instanton part. Let us introduce instanton expansion
parameter $q=\Lambda^4$ and expand $\mathcal{F}(a)$ and $u(a)$ in
terms of $k$-instanton contributions,\beq -\mathcal
{F}^{inst}&=&\sum_{k=1}^{\infty}\mathcal {F}_kq^k,\\
2u(a)&=&a^2+\sum_{k=1}^{\infty}\mathcal {G}_kq^k.\eeq Therefore we
have $\mathcal {F}_k=(\f{1}{i\pi})c_ka^{2-4k}$, and the equation
(\ref{elecrelat}) leads to the Matone's
relation\cite{matone1}\footnote{Our notations use a different
normalization compared to \cite{matone1}. We have absorbed a $4\pi
i$ factor into $\mathcal {F}_k$, and $\mathcal {G}_k$ is the
expansion for $2u$ compared to that in \cite{matone1} for $u$,
therefore there is a $2\pi i$ difference compared to (14) of
\cite{matone1}.} \be \mathcal {G}_k=k\mathcal
{F}_k.\label{elecmatone}\ee Equation (\ref{elecrelat}) and
(\ref{elecmatone}) are equivalent for Seiberg-Witten theory, however
as we will see later, for $\Omega$ deformed theory equation
(\ref{elecmatone}) is preserved while (\ref{elecrelat}) is not.

This relation was first found in \cite{matone1} for SU(2) theory
without matter by direct using Seiberg-Witten data. Generalization
to higher rank gauge groups were carried out in \cite{matonegneral}.
In the instanton calculation frame work this relation was proved in
\cite{matone1.5} up to two instantons, in \cite{matone3} for
muti-instanton contributions. It was also proved in \cite{wardid}
through the superconformal Ward identities of super Yang-Mills
applied to $\mathcal {N}=2$ gauge theory with broken gauge symmetry.

In fact, the prepotential depends on the scale $\Lambda$ in the form
$\mathcal {F}(a,\Lambda)=a^2f(\f{\Lambda}{a})$, the left hand side
of equation (\ref{elecrelat}) can be interpreted as the
renormalization flow of the prepotential, \be
2\mathcal{F}(a)-a\f{\p\mathcal{F}(a)}{\p
a}=\Lambda\f{\p\mathcal{F}(a)}{\p\Lambda}.\label{renoreq}\ee The
renormalization flow therefore incorporates the instanton effects.
This renormalization interpretation is not only limited to the pure
gauge theory\cite{BM96}, it was also derived in \cite{HKP96} for
SU(N) theory with massive fundamental matter, in \cite{HP97} for
SU(N) theory with adjoint matter, and in \cite{egm99} for general
gauge groups with matter.

An even more interesting fact is that the Matone's relation is a
result of integrability of Seiberg-Witten theory \cite{integsys,
integsys2,Whithamhiera, Whithamhiera2}. In a simpler form, the
integrability of Seiberg-Witten theory can be understood as a
precise correspondence between solutions of Coulomb branch of
various $\mathcal {N}=2$ theories and certain classical integrable
systems. For example, for the SU(N) pure $\mathcal {N}=2$ gauge
theory, the corresponding integrable system is the N-particle
periodic Toda chain. On a deeper level, the $\mathcal {N}=2$
theory/integrable models can be embedded into the Whitham-Toda
theory of integrable hierarchy \cite{integsys, Whithamhiera,
Whithamhiera2}. The Whitham-Toda system consists of a spectral curve
equipped with a meromorphic differential $dS$ depending on the
modulated moduli $h_k(T_n)$ and infinite number of ``slow" time
variables $T_n, n\ge1$(compared to the ``fast" time variables which
govern the evolution of high frequency oscillations, after averaging
these fast modes we get the Whitham dynamics of remaining slow modes
\cite{slowfast, slowfast2}). With these data we can construct period
integrals of $dS$ on the spectral curve, its $\tau$-function, and a
whole family of flow equations for $\tau$-function with respect to
slow times. The Seiberg-Witten data, including (\ref{elecrelat}) and
(\ref{renoreq}), can be obtained by proper parameters truncation and
redefinition.

The integrable hierarchy has a gauge theory realization as
``extended Seiberg-Witten theory" \cite{Whithamhiera2,
extendedhierach, extendedhierach2, instcount2, instcount4}, by
turning on higher Casimirs in the action,\be \mathcal {L}=\mathcal
{L}_0+\sum_{m>1}^{\infty}t_m\m{Tr}\Phi^m,\ee with nilpotent
couplings $t_m$, $\Phi$ is the $\mathcal {N}=2$ vector superfield.
Note that $t_2$ is actually a shift of complex gauge coupling
$\tau_0$ because the undeformed Lagrangian is $\mathcal
{L}_0=\tau_0\m{Tr}\Phi^2$. The effective prepotential now depend on
time variables $t_m$, denoted as $\mathcal {F}(a,\vec{t},\Lambda)$.
There exists an explicit correspondence between the extended
Seiberg-Witten theory and integrable hierarchy, if properly identify
variables on two sides. The scalar $a_i$ and $a_D^i$ is related to
the periods $\alpha_i=\oint_{A_i}dS$ and $\alpha_D^i=\oint_{B_i}dS$
of Whitham-Toda, and the prepotential $\mathcal{F}$ is related to
the logarithm of quasiclassical $\tau$-function of Whitham-Toda.

The Seiberg-Witten theory/Toda chain data can be recovered by
identifying $T_1$ with $\Lambda$ and setting $T_n=0, n\ge2$. The
equation (\ref{renoreq}) comes from the fact that the prepotential
$\mathcal {F}(\alpha_i,T_n)$ is a homogeneous function of degree
two,\be 2\mathcal {F}=\sum_{i}\alpha_i\f{\p \mathcal {F}}{\p
\alpha_i}+\sum_{n}T_n\f{\p \mathcal {F}}{\p T_n}.\ee And equation
(\ref{elecrelat}) comes from the above equation and the following
first order derivatives equation of prepotential,\be \f{\p \mathcal
{F}}{\p T_n}=\f{N}{i\pi n}\sum_{m}mT_mH_{m+1,n+1}.\ee We have to use
the fact $H_{2,2}=H_2=u_2=<\m{Tr}\varphi^2>$. The Whitham-Toda
hierarchy also contains equations of higher order derivatives of
prepotential with respect to slow times. Especially, explicit
results of the second order derivatives have been worked out in
\cite{extendedhierach2} through studying the higher Casimirs
deformed microscopic field theory, the equations are related to the
theta functions on the spectral curve, and the contact terms are
fixed. It is found that from the first and second order derivative
equations we can uniquely determine the prepotential of $\mathcal
{N}=2$ gauge theory in both weakly coupled region and strongly
coupled region\cite{Whithamhiera2, whithamstrong}.

As the last piece of background introduction, we would like to
mention that for theory with higher rank gauge groups, the third
order derivatives of prepotential with respect to $a_i$ satisfy the
generalized Witten-Dijkgraaf-Verlinde-Verlinde(WDVV) equations
\cite{wdvv}. This is also related to the theory of Whitham
hierarchies and topological field theories.

The structure of rest part of the paper is as follows. In the second
section we present the Matone's relation for gauge theory in the
$\Omega$ background with $\epsilon_1\ne0, \epsilon_2=0$, which does
not rely on the localization method. We show that the magnetic
expansion also satisfies the Matone's relation, if nonperturbative
monopole corrections are properly identified. This observation is
based on our previous work, and is the main reason for the present
work. In the third section we briefly review a general proof of
Matone's relation using localization technique. The last section is
devoted to discussion on the spectrum of periodic Toda chain in the
Nekrasov-Shatashvili scheme, Matone's relation indicates a relation
between the quantized energy and the scattering data of Toda
particles.

\section{Matone's relation in presence of $\Omega$-deformation: a case study}

Integrating multi-instanton contribution turns out to be very hard,
only first few multi-instanton results have been worked out and they
are in accord with Seinerg-Witten result. A remarkable development
in the last decade is the direct integration of instanton
contributions using the localization technique\cite{instcount,
instcount2, instcount3, instcount4}.  The ADHM construction of SU(N)
instanton solution gives the moduli space for general $k$-instanton
configurations $\mathcal {M}_{k,N}$ of $4kN$ dimension, which is
rotated by the global symmetry $SU(N)\times SO(4)$. The maximal
abelian subgroup is $U(1)^{N-1}\times U(1)^2$, generated by
$T_a=\m{exp}i\small{\m{diag}}(a_1,a_2,\cdots,a_N)$ constrained by
$\sum_i a_i=0$, and by
$T_{\pm}=\m{exp}i(\f{1}{2}\epsilon_{\pm}\sigma_3)$. $U(1)^{N-1}$ is
the maximal abelian subgroup of gauge group SU(N), $a_i$ is the
components of the v.e.v of the scalar field in the Coulomb phase;
$U(1)^2$ is the rotation of spacetime $\mathbb{R}^4$ along two
orthogonal planes $z_1,z_2$,
$\epsilon_{\pm}=\epsilon_1\pm\epsilon_2$ are complex parameters. The
idea is to use the $U(1)^{N-1}\times U(1)^2$ action of the
associated equivariant bundle $\mathcal{E}$ on $\mathcal
{M}_{k,N}\times\mathbb{C}^2$ to deform the gauge theory and
construct multi-instanton action that is equivalent closed form
under the group action. With this choice the group action has finite
number of isolated fixed points in the moduli space, we can apply
the localization formula to do integrals over $\mathcal {M}_{k,N}$.
This construction is equivalent to turning on a particular form of
background supergravity field parametered by $\epsilon_{1,2}$,
called $\Omega$-background \cite{instcount0, instcount, instcount2}.

The equivalent localization reduces the integrals to summation on
contribution from fixed points in the moduli space, results in the
Nekrasov's instanton partition function \cite{instcount}, a
holomorphic function depending on $a_i, \epsilon_1,\epsilon_2,
\Lambda$, and masses of hypermultiplets if presented, \be
Z=\m{exp}\f{\mathcal{F}(a,
\epsilon_1,\epsilon_2,\Lambda)}{\epsilon_1\epsilon_2},\ee where
$\f{1}{\epsilon_1\epsilon_2}$ is the leading singularity and
$\mathcal{F}(a, \epsilon_1,\epsilon_2, \Lambda)$ is a regular
function of $\epsilon_{1,2}$. The $\Omega$ deformed prepotential
again can be divided into perturbative part and nonperturbative
part, $\mathcal{F}(a, \epsilon_1,\epsilon_2)=\mathcal{F}^{pert}(a,
\epsilon_1,\epsilon_2)+\mathcal{F}^{inst}(a,
\epsilon_1,\epsilon_2)$. The Seiberg-Witten theory is recoved in the
limit $\epsilon_{1,2}\to 0$. Without causing confusion, we will use
the same symbols as in the Seiberg-Witten theory to express
quantities like $\mathcal {F}, a, u(a)$ of the $\Omega$ deformed
theory. We have to bear in mind that these quantities are all
promoted to depend on the deformation parameters $\epsilon_{1,2}$.

The $\Omega$ background is a delicate deformation of $\mathcal
{N}=2$ gauge theories. It greatly simplifies the instanton
integrals, meanwhile there are evidence that this deformation
preserves properties of the Seiberg-Witten theory, including
electric-magnetic duality and integrability. As it seems that the
deformation is just right, we expect the Matone's relation should
also be preserved, too. In fact, using localization technique, the
Matone's relation has been proved in the presence of $\Omega$
background\cite{matone2}. However, a careful identification of
$u(a)$ and the expectation of a scalar field is needed. We will
explain this in the next section, before going to that, let us do a
case study for a particular case which dose not rely on localization
technique .

Consider SU(N) pure super Yang-Mills theory in the $\Omega$
background with $\epsilon_1\ne0, \epsilon_2=0$. According to
\cite{NS0908}, the gauge theory in this background is identical to
the quantization of $A_{N-1}$ periodic Toda chain. In \cite{MM09} it
was proposed that the $\epsilon_1$ deformed partition function can
be obtained from the Bohr-Sommerfeld integrals of corresponding
quantum mechanical system. Along this line, in \cite{He2} we have
obtained the partition function of the SU(2) theory through the WKB
analysis of the cosine-potential quantum mechanics. Therefore, the
result we present in this section does not rely on localization
method and provides an independent evidence for the Matone's
relation\footnote{We notice in recent papers\cite{mt1006, po1006}
the $\epsilon_1$ deformed Matone's relation is discussed. Paper
\cite{mt1006} involves a semiclassical limit of Liouville CFT that
leads to quantum integrable system of \cite{NS0908}, the Matone's
relation is obtained from differential equations satisfied by the
conformal blocks with degenerate operators, theories with matter
hypermultiplets are discussed. Paper\cite{po1006} discusses
instanton counting when $\epsilon_2=0$ in the spirit
of\cite{instcount2}, general operators $\m{Tr}\phi^m$ are also
discussed in that formalism.}.

The deformation parameter $\epsilon_1$ has mass dimension one, when
viewed as a component of graviphoton field. The full prepotential
$\mathcal{F}(a,\epsilon_1,\Lambda)=\mathcal{F}^{pert}+\mathcal{F}^{inst}$
has dimension two, it can be written in the form $\mathcal
{F}=a^2f(\f{\Lambda}{a})g(\f{\epsilon_1}{a})$, with $f(x)$ contains
logarithms for perturbative effects and polynomials for instanton
corrections, and $g(x)$ contains only polynomials of even order.
Therefore, eq. (\ref{renoreq}) is modified to\be 2\mathcal
{F}-a\f{\p \mathcal {F}}{\p a}=\Lambda\f{\p\mathcal
{F}}{\p\Lambda}+\epsilon_1\f{\p\mathcal {F}}{\p\epsilon_1}.\ee This
can be easily checked using results of \cite{He2}. The
renormalization flow equation
$\Lambda\p_{\Lambda}\mathcal{F}=\f{2u}{i\pi}$ is not satisfied
anymore. Instead, we find\be \Lambda\f{\p \mathcal{F}}{\p
\Lambda}=\f{2u-\f{1}{24}\epsilon_1^2}{i\pi}.\ee The anomaly term
$\f{1}{24}\epsilon_1^2$ comes from the 1-loop perturbative
correction $\f{1}{24}\epsilon_1^2\m{ln}\f{a}{\Lambda}$. Clearly, the
structure of eq. (\ref{elecrelat}) is destroyed by $\epsilon_1$
corrections.

As the anomaly comes only from perturbative effects, we expect the
instanton corrections relation eq. (\ref{elecmatone}) is preserved.
We show that the instanton expansion indeed satisfies the Matone's
relation. Up to the first few instanton expansion and $\epsilon_1$
expansion, we have\beq
-\mathcal{F}^{inst}&=&\f{1}{2}a^{-2}q+\f{5}{64}a^{-6}q^2+\f{3}{63}a^{-10}q^3+\cdots\nn\\
&\quad&+\f{1}{8}\epsilon_1^2a^{-4}q+\f{21}{128}\epsilon_1^2a^{-8}q^2+\f{55}{192}\epsilon_1^2a^{-12}q^3+\cdots\nn\\
&\quad&+\f{1}{32}\epsilon_1^4a^{-6}q+\f{219}{1024}\epsilon_1^4a^{-10}q^2+\f{1495}{1536}\epsilon_1^4a^{-14}q^3+\cdots\eeq
where $q=\Lambda^4$ is the instanton expansion parameter. Tree and
1-loop contributions are not included. Writing $\mathcal {F}^{inst}$
as instanton expansion \be
-\mathcal{F}^{inst}=\sum_{k=1}^{\infty}\mathcal {F}_kq^k ,\ee now
$\mathcal {F}_k$ should be functions of $a$ and $\epsilon_1$. We
get\beq \mathcal
{F}_1&=&\f{1}{2}a^{-2}+\f{1}{8}\epsilon_1^2a^{-4}+\f{1}{32}\epsilon_1^4a^{-6}+\cdots\nn\\
\mathcal
{F}_2&=&\f{5}{64}a^{-6}+\f{21}{128}\epsilon_1^2a^{-8}+\f{219}{1024}\epsilon_1^4a^{-10}+\cdots\nn\\
\mathcal
{F}_3&=&\f{3}{64}a^{-10}+\f{55}{192}\epsilon_1^2a^{-12}+\f{1495}{1536}\epsilon_1^4a^{-14}+\cdots\eeq

On the other hand, we also have the expectation value
$u=<\m{Tr}\varphi^2>$ in the Seiberg-Witten theory now promoted to
\beq
2u&=&a^2+\f{1}{2}a^{-2}q+\f{5}{32}a^{-6}q^2+\f{9}{64}a^{-10}q^3+\cdots\nn\\
&\quad&+\f{1}{8}\epsilon_1^2a^{-4}q+\f{21}{64}\epsilon_1^2a^{-8}q^2+\f{55}{64}\epsilon_1^2a^{-12}q^3+\cdots\nn\\
&\quad&+\f{1}{32}\epsilon_1^4a^{-6}q+\f{219}{512}\epsilon_1^4a^{-10}q^2+\f{1495}{512}\epsilon_1^4a^{-14}q^3+\cdots.\label{typeAu}\eeq
Define $\mathcal {G}_k$ through \be
2u(a)=a^2+\sum_{k=1}^{\infty}\mathcal {G}_kq^k. \ee Then we have
\beq \mathcal
{G}_1&=&\f{1}{2}a^{-2}+\f{1}{8}\epsilon_1^2a^{-4}+\f{1}{32}\epsilon_1^4a^{-6}+\cdots\nn\\
\mathcal
{G}_2&=&\f{5}{32}a^{-6}+\f{21}{64}\epsilon_1^2a^{-8}+\f{219}{512}\epsilon_1^4a^{-10}+\cdots\nn\\
\mathcal
{G}_3&=&\f{9}{64}a^{-10}+\f{55}{64}\epsilon_1^2a^{-12}+\f{1495}{512}\epsilon_1^4a^{-14}+\cdots\eeq

$\mathcal {F}_k$ and $\mathcal {G}_k$ satisfy the Matone's relation
\be \mathcal {G}_k=k\mathcal {F}_k.\ee

The sum of instanton correction is just the density of instanton gas
\cite{instcount2}, \be \rho=\sum_{k=1}^{\infty}\mathcal
{G}_kq^k\sim\m{exp}-\f{8\pi^2}{g^2}.\ee In the weak coupling theory,
$g<<1$, the dilute gas approximation applies. However, if the
coupling is strong, we have to work in the dual magnetic theory.

In \cite{He2} we have also obtained the dual magnetic and dynoic
expansions of the prepotential
$\mathcal{F}_{D/T}(a_{D/T},\epsilon_1,\Lambda)$. The magnetic and
dynoic expansions are actually mirror to each other, therefore, we
only discuss magnetic case. The magnetic prepotential also have a
structure that can be explained as perturbative part and
nonperturbative part. The nonperturbative degrees of freedom is the
massive hypermultiplets of the dual magnetic theory, or the
monopoles of the original electric theory.

In fact, the Seiberg-Witten part of magnetic(dyonic) expansion also
satisfy equations similar to (\ref{elecrelat}). Including the
perturbative part, they are\cite{He2} \beq \mathcal
{F}_D(a_D)&=&\f{1}{i\pi}[\f{\hat{a}_D^2}{2}\ln(-\f{\hat{a}_D}{2})+4\hat{a}_D-\f{3}{4}\hat{a}_D^{2}
+\f{1}{16}\hat{a}_D^{3}+\f{5}{512}\hat{a}_D^{4}+\f{11}{4096}\hat{a}_D^{5}+\cdots],
\\
\sigma&=&-2\hat{a}_D+\f{1}{4}\hat{a}_D^2+\f{1}{32}\hat{a}_D^3+\f{5}{512}\hat{a}_D^4+\cdots.\eeq
They satisfy a dual form of equation (\ref{elecrelat}), namely \be
2\mathcal {F}_D-\hat{a}_D\f{\p \mathcal {F}_D}{\p
\hat{a}_D}=-\f{2\sigma}{i\pi},\label{magerelat}\ee For the dyonic
case, it is \be 2\mathcal {F}_T-a_T\f{\p \mathcal {F}_T}{\p
\hat{a}_T}=\f{2\varpi}{i\pi}.\label{dyonrelat}\ee with
$\varpi=u+\Lambda^2$. The left hand side of equations
(\ref{magerelat}) and (\ref{dyonrelat}) can be written as
renormalization equations, too.

Due to the $\epsilon_1$ corrections, equations (\ref{magerelat}) and
(\ref{dyonrelat}) are not satisfied by the dual deformed
prepotential $\mathcal{F}_D(a_D,\epsilon_1,\Lambda)$. However, if we
properly identify noperturbative corrections, the Matone's relation
still holds. The nonperturbative part we identify is \beq \mathcal
{F}_D^{monop}&=&\f{1}{16}\hat{a}_D^{3}q_D+\f{5}{512}\hat{a}_D^{4}q_D^2+\f{11}{4096}\hat{a}_D^{5}q_D^3+\cdots\nn\\
&\quad&+\f{\epsilon_1^2}{2^5}(-\f{3}{2^3}\hat{a}_Dq_D-\f{17}{2^7}\hat{a}_D^2q_D^2
-\f{205}{2^{10}\times3}\hat{a}_D^3q_D^3+\cdots)\nn\\
&\quad&+\f{\epsilon_1^4}{2^{11}}(\f{135}{2^9}\hat{a}_Dq_D^3+\f{2943}{2^{13}}\hat{a}_D^{2}q_D^4+\cdots).\label{magFD}\eeq
We have introduced the monopole expansion parameter
$q_D=\f{1}{\Lambda}$. Again, perturbative contributions are not
included. Define the monopole expansion as \be \mathcal
{F}_D^{monop}=\sum_{k=1}^{\infty}\mathcal {F}^D_kq_D^k, \ee then we
have \beq \mathcal
{F}^D_1&=&\f{1}{16}\hat{a}_D^3-\f{3}{2^8}\epsilon_1^2\hat{a}_D,\nn\\
\mathcal
{F}^D_2&=&\f{5}{512}\hat{a}_D^4-\f{17}{2^{12}}\epsilon_1^2\hat{a}_D^2,\nn\\
\mathcal {F}^D_3&=&\f{11}{4096}\hat{a}_D^5-\f{205}{2^{15}\times
3}\epsilon_1^2\hat{a}_D^3+\f{135}{2^{20}}\epsilon_1^4\hat{a}_D.\eeq
Note that these polynomials have finite terms, because of the ``gap"
structure of the dual expansion of prepotential.

The proper local moduli coordinate near $u=\Lambda^2$ is
$\sigma=u-\Lambda^2$, we have \beq
\sigma&=&-2\hat{a}_Dq_D^{-1}+\f{1}{4}\hat{a}_D^2+\f{1}{32}\hat{a}_D^3q_D+\f{5}{512}\hat{a}_D^4q_D^2+\f{33}{8192}\hat{a}_D^5q_D^3+\cdots\nn\\
&\quad&+\f{\epsilon_1^2}{2^6}(-1-\f{3}{8}\hat{a}_Dq_D-\f{17}{64}\hat{a}_D^2q_D^2-\f{205}{1024}\hat{a}_D^3q_D^3+\cdots)\nn\\
&\quad&+\f{\epsilon_1^4}{2^{17}}(9q_D^2+\f{405}{16}\hat{a}_Dq_D^3+\f{2943}{64}\hat{a}_D^2q_D^4+\f{69001}{1024}\hat{a}_D^3q_D^5+\cdots).\label{typeBsigma}\eeq
Define $\mathcal {G}_k^D$ through \be
2\sigma=<2\sigma>_0+\sum_{k=1}^{\infty}\mathcal {G}^D_kq_D^k, \ee
where $<2\sigma>_0$ represents terms that cannot be interpreted as
monopole corrections. For example, the first monopole correction
should starts from terms of order $q_D$. Also, the first term in
$\epsilon_1^4$ expansion seems not a monopole contribution, actually
all the first terms in higher $\epsilon_1$ expansions should be
excluded. The validity of the dual Matone's relation presented in
the following indicates validity of this chioce. We do not face this
problem for electric expansions since it is straightforward to
recognize all instanton terms. We have \beq \mathcal
{G}^D_1&=&\f{1}{16}\hat{a}_D^3-\f{3}{2^8}\epsilon_1^2\hat{a}_D,\nn\\
\mathcal
{G}^D_2&=&\f{5}{256}\hat{a}_D^4-\f{17}{2^{11}}\epsilon_1^2\hat{a}_D^2,\nn\\
\mathcal
{G}^D_3&=&\f{33}{4096}\hat{a}_D^5-\f{205}{2^{15}}\epsilon_1^2\hat{a}_D^3+\f{405}{2^{20}}\epsilon_1^4\hat{a}_D.\eeq

$\mathcal {F}^D_k$ and $\mathcal {G}^D_k$ have finite terms, they
also satisfy the Matone's relation \be \mathcal {G}^D_k=k\mathcal
{F}^D_k. \ee

The sum of monopole corrections is \be \rho_D=\sum_{k=1}^{\infty}
\mathcal {G}^D_kq_D^k\sim\m{exp}-\f{4\pi^2}{g_{\small{D}}^2},\ee
when the dual coupling $g_D$ is small, this can be interpreted as
density of dilute monopole gas.

The existence of Matone's relation in the magnetic theory can be
viewed as the consequence of compatibility of integrability and
duality. In fact, the associated integrable hierarchies are closely
related to modular forms on the spectral curve, as the $\tau$
function of algebraic integrable system is essentially the Riemann
theta function. Explicit results of the genus zero and genus one
gravitational contributions to the prepotential $\mathcal
{F}(a,\epsilon_1,\epsilon_2,\Lambda)$ presented in \cite{instcount4}
indicate their dependence on modular forms.

A bonus of the Matone's relation is application to the theory of the
Mathieu differential equation. We have analysed the relation between
SU(2) Seiberg-Witten gauge theory and the Mathieu equation in
\cite{He2}. The eigenvalue formulae (24) and (31) in the second
paper of \cite{He2} are precisely the instanton expansions, in the
terminology of gauge theory, if we notice the correspondence
$\lambda_\nu=8\epsilon_1^{-2}u,\nu=2\epsilon_1^{-1}a,
q=4\epsilon_1^{-2}\Lambda^2$ in that paper(Attention: these are the
symbols we have used in our previous paper \cite{He2}. The expansion
parameter $q$ is not the same $q=\Lambda^4$ in this paper, this
notation switch applies only when we refer to the Mathieu equation).
In fact, every coefficient of $q^{2k}$ of the Mathieu eigenvalue
expansion $\lambda_\nu$ in paper \cite{He2} is precisely $4\mathcal
{G}_k$ of $\m{Tr}\varphi^2$ of gauge theory. Indeed, it is easy to
check that the coefficients $\mathcal {F}_k$ obtained by instanton
counting, after setting $\epsilon_2=0$ and multiplying $4k$, exactly
coincide with formulae (24) and (31). For example, the first paper
of \cite{zk} gives explicit results of $\mathcal {F}_k$ up to
4-instantons convenient for our comparison, the first 3-instanton
corrections are \be
\mathcal{F}^{inst}|_{\epsilon_2=0}=\f{2}{4a^2-\epsilon_1^2}\Lambda^4+
\f{20a^2+7\epsilon_1^2}{(4a^2-\epsilon_1^2)^3(4a^2-4\epsilon_1^2)}\Lambda^8+
\f{16(144a^4+29\epsilon_1^4+232a^2\epsilon_1^2)}{3(4a^2-\epsilon_1^2)^5(4a^2-4\epsilon_1^2)(4a^2-9\epsilon_1^2)}\Lambda^{12}+\cdots.\ee
While the first 3-terms $q^2$-expansion of the Mathieu eigenvalue
$\lambda_\nu$ are \be
\lambda_\nu-\nu^2=\f{1}{2(\nu^2-1)}q^2+\f{5\nu^2+7}{32(\nu^2-1)^3(\nu^2-4)}q^4
+\f{9\nu^4+58\nu^2+29}{64(\nu^2-1)^5(\nu^2-4)(\nu^2-9)}q^6+\cdots.\ee
Precisely we have
$4k\mathcal{F}_k|_{\epsilon_2=0}=\lambda_\nu^{(k)}$, as expected.
Therefore, along the line of \cite{instcount} by counting all
possible partitions $k=k_1+k_2$ with the corresponding 2-Young
diagrams $\lbrace Y_1, Y_2\rbrace$, we can evaluate $\mathcal
{F}_k$, thus get $\mathcal {G}_k$ via the Matone's relation, which
essentially is the coefficient of $q^{2k}$ of the Mathieu eigenvalue
expansion for $\f{q}{\nu^2}<<1$.

However, the story of the magnetic theory is less well understood,
compare to the electric theory, although the dual relation is
observed. Notice that the formula (34) in the second paper of
\cite{He2} is \be \lambda_\nu=
2q-4\nu\sqrt{q}+\f{4\nu^2-1}{2^3}+\f{4\nu^3-3\nu}{2^6\sqrt{q}}
+\f{80\nu^4-136\nu^2+9}{2^{12}q}+\f{528\nu^5-1640\nu^3+405\nu}{2^{16}q^{\f{3}{2}}}+\cdots\ee
Terms of order $q^{\f{k}{2}}$ are precisely monopole corrections,
but few leading terms are not. Setting $\nu=0$ we get the ground
state energy $<\sigma>_0$, this includes terms such as
$9/(2^{12}q)$, which have the form of monopole corrections but are
incompatible with the dual prepotential expansion and dual Matone's
relation, thus excluded from $\mathcal {G}_k^D$. As we do not have
an explicit algorithm to counting monopole correction $\mathcal
{F}_k^D$, thus there is no Young-diagram-counting method to evaluate
the Mathieu eigenvalue expansion for $\f{q}{\nu^2}>>1$.

\section{Matone's relation from localization}

In \cite{matone2} it was proved that the Matone's relation
(\ref{elecmatone}) should hold for general $\epsilon_1,\epsilon_2$,
based on localization technique. We briefly review the results here
for completeness. We need to know instanton corrections to $\mathcal
{F}(a)$ and $u(a)$ of the $\Omega$ deformed theory. The prepotential
can be extracted from the Nekrasov's partition function. The
integral of $k$-instanton contribution localized to the fixed points
of the $U(1)^{N-1}\times U(1)^2$ action. These fixed points are in
one to one correspondence with the partitions of $N$ integers
$k_\alpha$ satisfying $k=k_1+k_2+\cdots+k_N$. Each $k_\alpha$ is
accompanied with a Young diagram $Y_\alpha$ with $k_\alpha$ boxes,
i.e. $|Y_\alpha|=k_\alpha$, therefor for each partition we have a
group of N-Young diagrams $\lbrace Y_\alpha, \alpha=1,2,\cdots
N\rbrace$. For each box $s$ in the $i_\alpha$-th row and
$j_\alpha$-th column of $Y_\alpha$, define the following counting
data: \be h_\beta(s)=\nu_{\beta,i_\alpha}-j_\alpha,\qquad
v_\beta(s)=\bar{\nu}_{\beta,j_\alpha}-i_\alpha.\ee where
$\nu_{\beta,i_\alpha}$ is the number of boxes in the $i_\alpha$-th
row of $Y_\beta$, and $\bar{\nu}_{\beta,j_\alpha}$ is the number of
boxes in the $j_\alpha$-th column of $Y_\beta$. Indexes $\alpha$ and
$\beta$ are not necessary the same, therefore when they are
different the box $s(i_\alpha,j_\alpha)$ may be located outside the
Young diagram $Y_\beta$. The quantity $h_\beta(s)$ is the horizontal
distance from the position of the box $s$ till the right end of the
$Y_\alpha$ diagram, and $v_\alpha(s)$ is the vertical distance from
the position of the box $s$ till the upper end of the $Y_\beta$
diagram.

Then the result of $k$-instanton contribution can be written in the
form \cite{zk}\be Z_k=\sum_{\lbrace
Y_\alpha,\sum|Y_\alpha|=k\rbrace}\prod_{\alpha,\beta=1}^N\prod_{s\in
Y_\alpha}\prod_{s^{'}\in Y_\beta}
\f{1}{E_{\alpha\beta}(s)(\epsilon_+-E_{\beta\alpha}(s^{'}))},\ee the
sum is running over all possible partitions, and\be
E_{\alpha\beta}(s)=a_{\alpha\beta}-h_\beta(s)\epsilon_1+(v_\alpha(s)+1)\epsilon_2,\ee
where $a_{\alpha\beta}=a_\alpha-a_\beta$. The full instanton
partition function is \be Z^{inst}=\sum_{k=1}^\infty Z_kq^k.\ee

For $u(a)$, we have to properly deform $<\m{Tr}\varphi^2>$ in the
Seiberg-Witten theory to suit the $\Omega$ background. In order to
use localization, we have to construct an integral
$<\m{Tr}\tilde{\varphi}^2>$ that is equivalent closed form under the
equivalent differential. Moreover, the integral must recover the
corresponding integral of the Seiberg-Witten $<\m{Tr}\varphi^2>$
when taking the limit $\epsilon_{1,2}\to0$. It turns out that the
proper deformed scalar field is \be
\tilde{\varphi}=\tilde{\varphi}_{bos}+\varphi_{ferm},\ee with \be
\tilde{\varphi}_{bos}=\bar{U}(z)i\left(\begin{array}{cc}
a & 0\\
0& \phi+\f{1}{2}\epsilon_-\sigma_3
\end{array}\right)U(z)-\bar{U}(z)iz_l\epsilon_l\f{\p}{\p z_l}U(z),
\ee where $U(z)$ comes from the self-dual gauge field
$A_\mu=\bar{U}(x)\p_\mu U(x)$. $\phi$ comes from
$T_{\phi}=\m{exp}i\phi\in U(k)$ which is the redundant symmetry of
the ADHM constraints, the value of its diagonal components $\phi_s$
at a fixed point are determined by the $k$-boxes of the
corresponding Young diagrams $\lbrace Y_\alpha\rbrace$,\be
\phi_{(i_\alpha,j_\alpha)}=a_\alpha+(j_\alpha-1)\epsilon_1+(i_\alpha-1)\epsilon_2.\ee
$\varphi_{ferm}$ is the solution of the scalar field when no v.e.v.
and $\epsilon$ deformation presented, constructed from ADHM data,
involving a fermion bilinear therefore the subscript.

Actually, using the localization technique a general formula for
chiral ring operators $<\m{Tr}\tilde{\varphi}^m>$ can be obtained,
it is \cite{extendedhierach, instcount4,  matone2} \be
<\m{Tr}\tilde{\varphi}^m>=\f{1}{Z^{inst}}\sum_{k=1}^\infty\sum_{\lbrace
Y_\alpha,\sum|Y_\alpha|=k\rbrace}\f{\mathcal{O}_m(\lbrace
Y_\alpha\rbrace)}{\prod_{\alpha,\beta=1}^N\prod_{s\in
Y_\alpha}\prod_{s^{'}\in
Y_\beta}E_{\alpha\beta}(s)(\epsilon_+-E_{\beta\alpha}(s^{'}))}q^k.\ee
with \beq \mathcal{O}_m(\lbrace
Y_\alpha\rbrace)&=&\sum_{\alpha=1}^N\lbrace
a_\alpha^m-\sum_{s(i_\alpha,j_\alpha)\in Y_\alpha}\lbrack
(a_\alpha+j_\alpha\epsilon_1+i_\alpha\epsilon_2)^m-(a_\alpha+j_\alpha\epsilon_1+(i_\alpha-1)\epsilon_2)^m\nn\\
&\quad&-(a_\alpha+(j_\alpha-1)\epsilon_1+i_\alpha\epsilon_2)^m
+(a_\alpha+(j_\alpha-1)\epsilon_1+(i_\alpha-1)\epsilon_2)^m\rbrack\rbrace.\eeq
can be read from the Chern character of the equivariant bundle
$\mathcal {E}$ over a fixed point, \be Ch_{\vec{Y}}(\mathcal
{E})=\sum_{\alpha}^N[e^{ia_\alpha}-(1-e^{i\epsilon_1})(1-e^{i\epsilon_2})\sum_{s(i_\alpha,j_\alpha)\in
Y_\alpha}e^{i[a_\alpha+(j_\alpha-1)\epsilon_1+(i_\alpha-1)\epsilon_2]}].\ee

For the case of $m=2$, the operator $\mathcal{O}_2(\lbrace
Y_\alpha\rbrace)$ greatly simplifies and does not depend on the
shape of Young diagram, \beq \mathcal{O}_2(\lbrace
Y_\alpha\rbrace)&=&\sum_{\alpha=1}^N(a_\alpha^2-2k_\alpha\epsilon_1\epsilon_2)\nn\\
&=&a_1^2+a_2^2+\cdots a_N^2-2k\epsilon_1\epsilon_2.\eeq Therefore we
have \beq <\m{Tr}\tilde{\varphi}^2>&=&\f{1}{\sum_k
Z_kq^k}\sum_k\sum_{\lbrace
Y_\alpha\rbrace}\prod_{\alpha,\beta=1}^N\prod_{s\in
Y_\alpha}\prod_{s^{'}\in
Y_\beta}\f{(\sum_\gamma a_\gamma^2-2k_\gamma\epsilon_1\epsilon_2)q^k}{E_{\alpha\beta}(s)(\epsilon_+-E_{\beta\alpha}(s^{'}))}\nn\\
&=&\sum_\gamma a_\gamma^2-2\epsilon_1\epsilon_2\f{\sum_kZ_kkq^k}{Z^{inst}}\nn\\
&=&\sum_\gamma a_\gamma^2-2\epsilon_1\epsilon_2q\f{\p}{\p
q}\m{ln}Z^{inst}\nn\\
&=&a_1^2+a_2^2+\cdots+a_N^2+\sum_{k=1}^\infty
2k\mathcal{F}_kq^k.\eeq We have used
$\epsilon_1\epsilon_2\m{ln}Z^{inst}=\mathcal{F}^{inst}=-\sum_k\mathcal{F}_kq^k$.
Considering $u(a,\epsilon_1,\epsilon_2,
\Lambda)=<\m{Tr}\tilde{\varphi}^2>$ can also be expanded as
$u(a)=a_1^2+a_2^2+\cdots+a_N^2+\sum_{k=1}^\infty 2\mathcal{G}_kq^k$
by definition, then the Matone's relation
$\mathcal{G}_k=k\mathcal{F}_k$ immediately follows.

\section{Implication for energy spectrum of Toda chain}

According to Nekrasov-Shatashvili scheme\cite{NS0908}, a large class
of four dimensional $\mathcal{N}=2$ super Yang-Mills theories
deformed by $\Omega$ background with $\epsilon_1\ne0, \epsilon_2=0$
are identical to quantization of the corresponding classical
integrable systems \cite{integsys, integsys2}. The deformation
parameter $\epsilon_1$ plays the role of the Plank constant.

The integrable model/$\mathcal {N}=2$ theory correspondence
establishes a relation between the periodic Toda chain and $\mathcal
{N}=2$ pure super Yang-Mills. The periodic Toda chain of length $L$
is defined as, in a dimensionless form,\be
H=\f{1}{2}\sum_{i=1}^{N}p_i^2+e^{-\f{L}{N}}\sum_{i=1}^{N}e^{x_i-x_{i+1}},\ee
with periodic boundary condition $x_{N+1}=x_1$. Its stationary
region is at $x_1=x_2=\cdots=x_N$. The open Toda chain can be
obtained by cutting the closed chain, i.e., eliminating the
interaction term $\m{exp}(x_N-x_1)$, its asymptotic region is at
$x_i-x_{i+1}\to-\infty$.

Gutzwiller studied the quantization condition for periodic Toda
chain in \cite{Gutzwiller}, based on the work of Kac and van
Moerbeke on classical chain \cite{kvm}. Kac and van Moerbeke
constructed a new set of conjugate coordinates variables for
N-particle periodic chain form the asymptotic momenta of N-1
particle open chain obtained by cutting the closed chain. This is a
nonlinear canonical transformation, the resulting mechanical system
gives the same conserved quantities as the original one, but have
simple evolution behavior: its equation of motion is separable and
breaks up into N-1 equations depending on only one variable.
Gutzwiller's method is the quantum analogy of the classical
transformation. The wave function of quantum N-particle periodic
chain $\Psi_N(x_1,x_2,\cdots x_N)$ is constructed as linear
combination of asymptotic wave function of N-1 particle open Toda
chain $\psi_{N-1}(x_1,x_2,\cdots x_{N-1};\lbrace k_i\rbrace)$. The
wave numbers $k_i$ are zeros of the Hill-type determinant. In the
case of $N=2$, this construction coincides with theory of modified
Mathieu differential equation. The vanishing of the wave function
for large separation $\m{exp}(x_i-x_{i+1})>>1$ requires $k_i$
satisfy the Gutzwiller's quantization conditions.  Essentially
Gutzwiller's solution of periodic Toda chain spectrum is in the
realm of Sklyanin's separation of variables\cite{sov}. However, the
resulting quantization conditions involve both zeros of Hill's
determinant and zeros of the characteristic polynomial of Lax
matrix, it is not easy to handle in practice.

The Nekrasov-Shatashvili scheme provides a novel way to quantize
these integrable systems. The main point is that the prepotnetial of
the gauge theory is identified with the Yang-Yang
function\cite{YY1969} of the integrable system, $a_i$ is identified
with the momentum of the $i$-th (quasi)particle, and the instanton
corrections are identified with finite size corrections
$\Lambda^{2N}\sim e^{-L}$. The quantization condition is given by
the Bethe equation, \be \f{\p \mathcal {F}(a,\epsilon_1,\Lambda)}{\p
a_i}=n_i, \qquad n_i\in \mathbb{Z}.\label{bethe}\ee As demonstrated
in \cite{KozlowskiTeschner}, the Bethe equation (\ref{bethe}) is
consistent with Gutzwiller's quantization of periodic Toda chain.
The eigenvalues of quantum Hamiltonians can be obtained from the
expectation values of chiral ring operators
$\m{Tr}\tilde{\varphi}^m$. Deform the gauge theory action as \be
\mathcal {L}_{\epsilon,t}=\mathcal
{L}_{\epsilon}+\sum_{m>1}t_m\m{Tr}\tilde{\varphi}^m, \ee where
$\mathcal {L}_{\epsilon}$ is the $\Omega$ deformed action. The
deformed action is still an equivalent closed form and the
localization can be applied, result in a prepotential $\mathcal
{F}(a,\epsilon_1,\epsilon_2,\vec{t},\Lambda)|_{\epsilon_2\to0}$.
Then we have \be \mathcal {E}_m=\f{\p\mathcal
{F}(a,\epsilon_1,\vec{t},\Lambda)}{\p t_m}|_{\vec{t}=0}.\ee

Therefore, using the explicit formulae presented in
\cite{extendedhierach, instcount4, matone2}, the quantum Hamiltonian
of periodic Toda chain can be evaluated, at least in the electric
frame. Especially, $\mathcal {E}_2$ is the energy of the Toda chain.
The Matone's relation discussed in the last two sections tells us
that $\mathcal {E}_2$ can be directly read from the Yang-Yang
function,\be \mathcal
{E}_2=a_1^2+a_2^2+\cdots+a_N^2+\sum_{k=1}^\infty
2k\mathcal{F}_kq^k|_{\epsilon_2=0}.\ee This is in accordance with
(6.13) of \cite{NS0908}. The momenta $a_i$ have to obey the
condition $a_1+a_2+\cdots+a_N=0$ and the Bethe quantization
condition (\ref{bethe}).

In the discussion above, $a_i$ are the momenta of the quasi
particles. In interacting multi particle systems quasiparticles are
the proper degrees of freedom(d.o.f) to describe the dynamics of
excitations. For example, the effective excitation d.o.f of spin
system are magnons, and the effective excitation d.o.f of crystal
lattice are phonons. When $a_i$ satisfy the large momenta condition
$a_i>>\Lambda, a_i-a_j>>\Lambda$, the gauge theory is weakly
coupled. It turns out that the properties of the excitations of
periodic Toda chain change when the momenta become small. This is
related to the electric-magnetic duality of the corresponding gauge
theory.

From the point of gauge theory this is quite reasonable, as the
electric-magnetic duality is main tool to understand dynamics of
gauge theories \cite{seibergdual}. The spirit is that at different
part in the moduli space, there exists different d.o.f that are
convenient for the formulation of the theory. In the weakly coupled
region, the proper d.o.f are the weakly coupled electric fields; in
the strongly coupled region, the proper d.o.f are the dual magnetic
fields. In the gauge theory/integrable system correspondence, the
moduli space of gauge theory is isomorphic to the space of
Hamiltonians of (complexified) integrable system. Therefore,
translate the logic of gauge theory duality to the corresponding
integrable system, it indicates the quasiparticle d.o.f may change
when we move from one region to another region in the space of
Hamiltonians $\lbrace\mathcal {E}_m\rbrace$. For the case of Toda
chain, if the momenta, therefor the Hamiltonians, are all large, the
proper d.o.f are the quasi particles with momenta $a_i$. The leading
order dispersion relation is nonrelativistic, $\mathcal {E}_2\sim
a^2$. These short wave length particles feel finite size corrections
scale as $\delta\mathcal {E}_2\sim (\f{\Lambda}{a})^{2Nk}$. However,
at certain point corresponding to the strongly coupled gauge theory,
the proper d.o.f should be some other kind quasiparticles with small
momenta denoted as $a_D$, satisfying $a_D<<\Lambda$. For example, at
the point of the maximal strong coupling singularities where the
maximal number of mutually local particles become massless, the
excitations of Toda chain are all magnetic quasiparticles with
momenta $a_{Di}$. This happens when the maximal number of distinct
pairs of branch points of the Seiberg-Witten curve(spectral curve)
collide, i.e. the curve
$y^2=P_N^2(x)-\Lambda^{2N}=\prod_{i=1}^{2N}(x-e_i(u_k))$ has only
double zeros. Each colliding pair of branch points corresponds to a
shrinking homology cycle, as the intersection number of these
shrinking homology cycles are zero, the corresponding massless
particles are mutually local. We have($u=u_2$)
\cite{DouglasShenker}\be
u-u_{max}=-2\Lambda\sum_{i=1}^{N-1}\hat{a}_{Di}\sin\f{i\pi}{N}+\f{1}{4}\sum_{i=1}^{N-1}\hat{a}_{Di}^2+\mathcal
{O}(\f{\hat{a}_{Di}^3}{\Lambda}).\ee  Other higher Casimirs
$u_m=<\m{Tr}\varphi^m>$ also have linear leading order dispersion
relations(see (19) of \cite{whithamstrong}). Therefor, for
quasiparticles of this kind the leading order dispersion relation is
relativistic, $\mathcal {E}_2\sim a_D$. These long wave length
quasiparticles feel finite size corrections scale as $\delta\mathcal
{E}_2\sim (\f{a_D}{\Lambda})^{k}$.

In order to have a full understanding about the quantum spectrum of
small momenta excitations, we have to work out the modular form of
the $\Omega$ deformed extended prepotential $\mathcal {F}(a,
\epsilon_1, \vec{t}, \Lambda)$ for general gauge groups, this is
still a challenging problem.

\section*{Acknowledgments}
W.H. thanks Prof. Bo Feng for valuable discussion.


\begin{thebibliography}{10}
\bibitem{SW9407}  N. Seiberg, E. Witten, ``Electric-Magnetic Duality, Monopole Condensation, And Confinement
In N = 2 Supersymmetric Yang-Mills Theory", Nucl.Phys. {\bf B426},
19-52(1994), Erratum-ibid. {\bf B430}, 485-486(1994),
[hep-th/9407087];\\``Monopoles, Duality and Chiral Symmetry Breaking
in N=2 Supersymmetric QCD", Nucl.Phys. {\bf B431}, 484-550(1994),
[hep-th/9408099].

\bibitem{matone1}
M. Matone, ``Instantons and recursion relations in N=2 Susy gauge
theory'',  Phys. Lett. {\bf B}357 (1995) 342, [hep-th/9506102].

\bibitem{matonegneral}
J. Sonnenschein, S. Theisen, S. Yankielowicz, ``On the Relation
Between the Holomorphic Prepotential and the Quantum Moduli in SUSY
Gauge Theories'',  Phys.Lett. B367 (1996) 145-150,
[hep-th/9510129].\\
T. Eguchi, S-K. Yang, ``Prepotentials of N=2 Supersymmetric Gauge
Theories and Soliton Equations", Mod.Phys.Lett. A11 (1996) 131-138,
[hep-th/9510183].

\bibitem{matone1.5}
F. Fucito, G. Travaglini, ``Instanton Calculus and Nonperturbative
Relations in N=2 Supersymmetric Gauge Theories'',      Phys.Rev.
{\bf D}55 (1997) 1099-1104, [hep-th/9605215].

\bibitem{matone3}
N. Dorey, V.V. Khoze, M.P. Mattis, ``Multi-Instanton Check of the
Relation Between the Prepotential $F$ and the Modulus $u$ in N=2
SUSY Yang-Mills Theory'',  Phys.Lett. B390 (1997) 205-209,
[hep-th/9606199].

\bibitem{wardid}
P. S. Howe, P. C. West, ``Superconformal Ward Identities and N=2
Yang-Mills Theory'',  Nucl.Phys. {\bf B} 486 (1997) 425-442,
[hep-th/9607239].

\bibitem{BM96}
G. Bonelli, M. Matone, ``Nonperturbative Renormalization Group
Equation and Beta Function in N=2 SUSY Yang-Mills", Phys.Rev.Lett.
76 (1996) 4107-4110, [hep-th/9602174].

\bibitem{HKP96}
Eric D'Hoker, I.M. Krichever, and D.H. Phong, ``The Renormalization
Group Equation in N=2 Supersymmetric Gauge Theories", Nucl.Phys.
B494 (1997) 89-104, [hep-th/9610156].

\bibitem{HP97}
Eric D'Hoker, D.H. Phong, ``Calogero-Moser Systems in SU(N)
Seiberg-Witten Theory", Nucl.Phys. B513 (1998) 405-444,
[hep-th/9709053].

\bibitem{egm99}
J. D. Edelstein, M. Gomez-Reino, J. Mas, ``Instanton corrections in
N=2 supersymmetric theories with classical gauge groups and
fundamental matter hypermultiplets", Nucl.Phys. {\bf B}561 (1999)
273-292, [hep-th/9904087].

\bibitem{integsys}
A. Gorsky, I.M. Krichever, A. Marshakov, A. Mironov and A. Morozov,
``Integrability and Seiberg-Witten exact solution" Phys. Lett. {\bf
B355}(1995)466 [hep-th/9505035]. \\
E. Martinec, N.Warner, ``Integrable systems and supersymmetric gauge
theories", Nucl. Phys. {\bf B}459 (1996) 97-112, [hep-th/9509161].
\bibitem{integsys2}
R. Donagi, E. Witten, ``Supersymmetric Yang-Mills and integrable
systems", Nucl.Phys. {\bf B}460(1996)299-334, [hep-th/9510101].

\bibitem{Whithamhiera}
T. Nakatsu, K. Takasaki, ``Whitham-Toda hierarchy and N = 2
supersymmetric Yang-Mills theory",     Mod.Phys.Lett. A11 (1996)
157-168, [hep-th/9509162].\\
H. Itoyama, A. Morozov, ``Prepotential and the Seiberg-Witten
Theory", Nucl.Phys. {\bf B}491 (1997) 529-573, [hep-th/9512161]

\bibitem{Whithamhiera2}
A. Gorsky, A. Marshakov, A. Mironov, A. Morozov, ``RG Equations from
Whitham Hierarchy", Nucl.Phys.B527(1998)690-716, [hep-th/9802007].\\
J. D. Edelstein, M. Marino, J. Mas, ``Whitham Hierarchies, Instanton
Corrections and Soft Supersymmetry Breaking in N=2 SU(N) Super
Yang-Mills Theory", Nucl.Phys. {\bf B}541 (1999) 671-697,
[hep-th/9805172].

\bibitem{slowfast}
I. M. Krichever, ``The $\tau$-function of the universal Whitham
hierarchy, matrix models and topological field theories", Commun.
PureAppl. Math.47(1994)437, [hep-th/9205110].

\bibitem{slowfast2}
O. Babelon, D. Bernard, M. Talon, {\em ``Introduction to Classical
Integrable Systems''}, Chapter 10, Cambridge University Press,
Cambridge, 2003.

\bibitem{whithamstrong}
J. D. Edelstein, J. Mas, ``Strong coupling expansion and
Seiberg-Witten-Whitham equations", Phys.Lett. {\bf B}452 (1999)
69-75, [hep-th/9901006].

\bibitem{extendedhierach}
A. Losev, A. Marshakov, N. Nekrasov, ``Small instantons, little
strings and free fermions", In ``From fields to strings", vol. {\bf
1}, 581-621, [hep-th/0302191].

\bibitem{extendedhierach2}
A. Marshakov, N. Nekrasov, ``Extended Seiberg-Witten Theory and
Integrable Hierarchy", JHEP 0701( 2007)104, [hep-th/0612019].


\bibitem{wdvv}
G. Bonelli, M. Matone, ``Nonperturbative Relations in N=2 SUSY
Yang-Mills and WDVV equation",     Phys.Rev.Lett. 77 (1996)
4712-4715, [hep-th/9605090].\\
 K. Ito, S-K. Yang, ``The WDVV
Equations in N=2 Supersymmetric Yang-Mills Theory", Phys.Lett. {\bf
B}433 (1998) 56-62,
[hep-th/9803126].\\
A.Marshakov, A.Mironov, A.Morozov, ``WDVV-like equations in N=2 SUSY
Yang-Mills Theory", Phys.Lett. {\bf B}389 (1996)43-52,
[hep-th/9607109]. ``More Evidence for the WDVV Equations in N=2 SUSY
Yang-Mills Theories",     Int.J.Mod.Phys. {\bf A}15 (2000)
1157-1206, [hep-th/9701123].

\bibitem{instcount0}
A. Losev, N. Nekrasov, S. L. Shatashvili, ``Issues in Topological
Gauge Theory",  Nucl. Phys. {\bf B} 534 (1998) 549,
[hep-th/9711108].\\
G. Moore, N. Nekrasov, S. L. Shatashvili, ``Integrating Over Higgs
Branches", Commun.Math.Phys. 209 (2000) 97-121, [hep-th/9712241].

\bibitem{instcount}
N. Nekrasov, ``Seiberg-Witten Prepotential From Instanton Counting",
Adv. Theor. Math. Phys. {\bf 7}, 831-864(2004), [hep-th/0206161].
Also, Proceedings of the ICM, Beijing 2002, vol. {\bf 3}, 477--496,
[hep-th/0306211]

\bibitem{instcount2}
N. Nekrasov and A. Okounkov, ``Seiberg-Witten Theory and Random
Partitions", [hep-th/0306238].

\bibitem{instcount3}
H. Nakajima, K. Yoshioka, ``Instanton counting on blowup. I.
4-dimensional pure gauge theory", Inventiones Mathematicae, Vol.162,
Number {\bf 2}, 313-355, [math/0306198].

\bibitem{instcount4}
H. Nakajima, K. Yoshioka, ``Lectures on Instanton Counting",
Proceedings of ``Workshop on algebraic structures and moduli
spaces", July 2003, [math/0311058].

\bibitem{matone2}
R. Flume, F. Fucito, J. F. Morales, R. Poghossian, ``Matone's
Relation in the Presence of Gravitational Couplings'',  JHEP 0404
(2004) 008, [hep-th/0403057].

\bibitem{zk}
R. Flume and R. Poghossian, ``An Algorithm for the Microscopic
Evaluation of the Coefficients of the Seiberg-Witten Prepotential",
Int. J. Mod. Phys. {\bf A} 18 (2003)
2541, [hep-th/0208176].\\
 U. Bruzzo, F. Fucito, J. F. Morales, A.
Tanzini, ``Multi-Instanton Calculus and Equivariant Cohomology'',
JHEP 0305(2003)054, [hep-th/0211108].

\bibitem{NS0908} N. Nekrasov, S. Shatashvili, ``Quantization of Integrable Systems and Four Dimensional Gauge Theories",
16th International Congress on Mathematical Physics, Prague, August
2009, P. Exner, Editor, pp.265-289, World Scientific 2010,
[arXiv:0908.4052].

\bibitem{MM09} A. Mironov, A. Morozov, ``Nekrasov Functions and Exact Bohr-Sommerfeld Integrals'',
JHEP. {\bf 04}(2010)040, [arXiv:0910.5670 [hep-th]]. ``Nekrasov
Functions from Exact BS Periods: the Case of SU(N)'', J.Phys. {\bf
A43}(2010)195401, [arXiv:0911.2396[hep-th]]. A. Popolitov, ``On
relation between Nekrasov functions and BS periods in pure SU(N)
case'', [arXiv:1001.1407[hep-th]].

\bibitem{He2}
W. He, Y-G. Miao, ``Magnetic expansion of Nekrasov theory: the SU(2)
pure gauge theory'', Phys. Rev.  {\bf D 82} (2010)025020,
[arXiv:1006.1214 [hep-th]]. ``Mathieu equation and Elliptic curve",
[arXiv:1006.5185[math-ph]].

\bibitem{mt1006} K. Maruyoshi, M. Taki, ``Deformed Prepotential, Quantum Integrable System and Liouville Field Theory'',
Nucl.Phys. {\bf B}841(2010)388-425, [arXiv:1006.4505 [hep-th]].

\bibitem{po1006} R. Poghossian, ``Deforming SW curve'', [arXiv:1006.4822 [hep-th]].

\bibitem{Gutzwiller}
M. C. Gutzwiller, ``The quantum mechanical Toda lattice'', Ann.
Phys. {\bf 124}(1980)347; ``The quantum mechanical Toda lattice
II'', Ann. Phys.{\bf 133}(1981)304.

\bibitem{kvm}
M. Kac, P. van Moerbeke, ``On Some Periodic Toda Lattices'', Proc.
Nat. Acad. Sci. USA {\bf 72}, No. 4(1975)1627-1629; ``A complete
solution of the periodic Toda problem'', Proc. Nat. Acad. Scd. USA
{\bf 72}, No. 8(1975)2879-2880.\\ P. van Moerbeke, ``The Spectrum of
Jacobi Matrices'', Inventiones math. {\bf 37}(1976)45-81.

\bibitem{sov}
E. K. Sklyanin, ``The quantum Toda chain'', Lecture Notes in
Physics, vol. {\bf 226}, Springer, Berlin, 1985; ``Separation of
Variables. New Trends.'', Prog. Theor. Phys. Suppl. 118(1995)35-60,
[arXiv:solv-int/9504001].

\bibitem{YY1969}C. N. Yang, C. P. Yang, ``Thermodynamics of 1-d system of Bosons with repulsive delta function interaction",
J. Math. Phys. {\bf 10} (1969), 1115.

\bibitem{KozlowskiTeschner}
K. K. Kozlowski, J. Teschner, ``TBA for the Toda
chain'', [arXiv:1006.2906 [math-ph]].

\bibitem{seibergdual}
N. Seiberg, ``Exact Results on the Space of Vacua of Four
Dimensional SUSY Gauge Theories", Phys.Rev. {\bf
D}49(1994)6857-6863, [hep-th/9402044].\\ K. Intriligator, N.
Seiberg, ``Lectures on supersymmetric gauge theories and
electric-magnetic duality'', Nucl.Phys.Proc.Suppl.45BC(1996)1-28,
[hep-th/9509066].

\bibitem{DouglasShenker}
M. R. Douglas and S. H. Shenker, ``Dynamics of SU(N) supersymmetric
gauge theory'', Nucl. Phys. {\bf B447}(1995)271-296,
[hep-th/9503163]. E. D'Hoker, D.H.Phong,``Strong coupling expansions
of SU(N) Seiberg-Witten theory'', Phys.Lett.{\bf
B397}(1997)94-103,[hep-th/9701055].



\end{thebibliography}
\end{document}